\documentclass[a4paper,12pt]{article}
\usepackage{graphicx}
\usepackage{color}
\begin{document}

\begin{center}
\textbf{\Large 
Entangling Time-Bin Qubits with a Switch
}
\vspace{2mm} 

$^*$Hiroki Takesue\\

\vspace{2mm}
{\it 

NTT Basic Research Laboratories, NTT Corporation, \\
3-1 Morinosato Wakamiya, Atsugi, Kanagawa, 243-0198, Japan\\

$^*$Corresponding author: takesue.hiroki@lab.ntt.co.jp

}

\end{center}






Abstract 

A time-bin qubit is a promising candidate qubit for advanced quantum information processing systems operating over optical fibers or integrated quantum photonic circuits. However, the single- and two-qubit operations of time-bin qubits have been largely unexplored compared with those of polarization qubits. Here, I present a simple scheme for implementing a basic two-qubit operation, namely an ``entangling" operation of two independent time-bin qubits, based on Hong-Ou-Mandel quantum interference in a high-speed 2x2 optical switch. With the proposed scheme, the formation of a time-bin entanglement was confirmed experimentally. 



\clearpage

Experimental quantum information processing (QIP) using photons has progressed significantly in the last decade. So far, most photonic QIP experiments have been performed using polarization qubits \cite{kok}. This is mainly because a state of polarization can be well preserved in free space, where the majority of such experiments were undertaken. In addition, single- and two-qubit gate operations are established with polarization qubits: i.e. most single- and two-qubit operations can be realized using simple optical elements such as beamsplitters, waveplates and polarization beam splitters (PBS). As a result, various QIP tasks such as quantum teleportation, quantum gates, and multi-photon entanglement generation have been demonstrated using polarization qubits and free space optics \cite{kok,pan}. 

However, if we go beyond proof-of-principle experiments and construct quantum information systems for real world applications, free space optics are not an appropriate choice in many cases. In fact, most long-distance terrestrial quantum communication experiments used optical fibers as quantum channels, where it is difficult to preserve the state of polarization because of birefringence. On the other hand, a new research field named ``integrated quantum photonics" has emerged recently \cite{plc,politi,peruzzo,crespi,broome,spring,tillmann}. The idea is to integrate quantum optical circuits using optical waveguide devices, so that we can realize large-scale photonic QIP systems with better phase stability and compactness. In this approach, too, most optical waveguides exhibit polarization dependence. For example, silicon waveguides, which have been used to realize functional devices such as photon sources \cite{sharping,silicon,matsuda} and buffers \cite{buffer} on a chip, generally have strong polarization dependent losses and dispersion. Silica waveguides usually have less polarization dependence, but it is still difficult to eliminate it completely. For example, the polarization dependent transmission spectra of a 1-bit delayed Mach-Zehnder interferometer fabricated on a silica waveguide is reported in \cite{honjo}. Although intensive researches are being undertaken to realize photonic quantum integrated circuits that supports polarization qubits \cite{sansoni,crespi}, it is unclear if polarization qubits are suitable for use in integrated quantum photonics.

In contrast, a time-bin qubit, which is a coherent superposition of single photon states in two or more different temporal modes, has been intensively used in quantum communication over optical fiber \cite{gisin,brendel}. Since this qubit possesses only a single polarization mode, it is naturally useful for the transmission of quantum information in optical fibers and waveguides. 
However, although there were a few multi-photon experiments based on time-bin entanglementsuch as quantum teleportation \cite{geneva} and entanglement swapping \cite{swp2,swp}, time-bin qubits have not yet been intensively used in sophisticated QIP experiments, mainly because single- and two-qubit operations have been largely unexplored with them. 
Although it is well known that a time-bin qubit can be converted into a polarization qubit using a simple linear optics circuit \cite{ppln,berlin}, such qubit conversion is accompanied by additional experimental errors and losses and is thus hard to implement in real systems. In this paper, I propose and demonstrate a simple scheme for implementing a basic two-qubit operation, namely an ``entangling" operation to two independent time-bin qubits using time-bin switching. The proposed scheme is useful for realizing various quantum gates such as a controlled-phase gate (CZ gate) \cite{langford} and fusion gates for cluster state quantum computation \cite{browne} for time-bin qubits.

\begin{figure}[thb]

\centerline{\includegraphics[width=\linewidth]{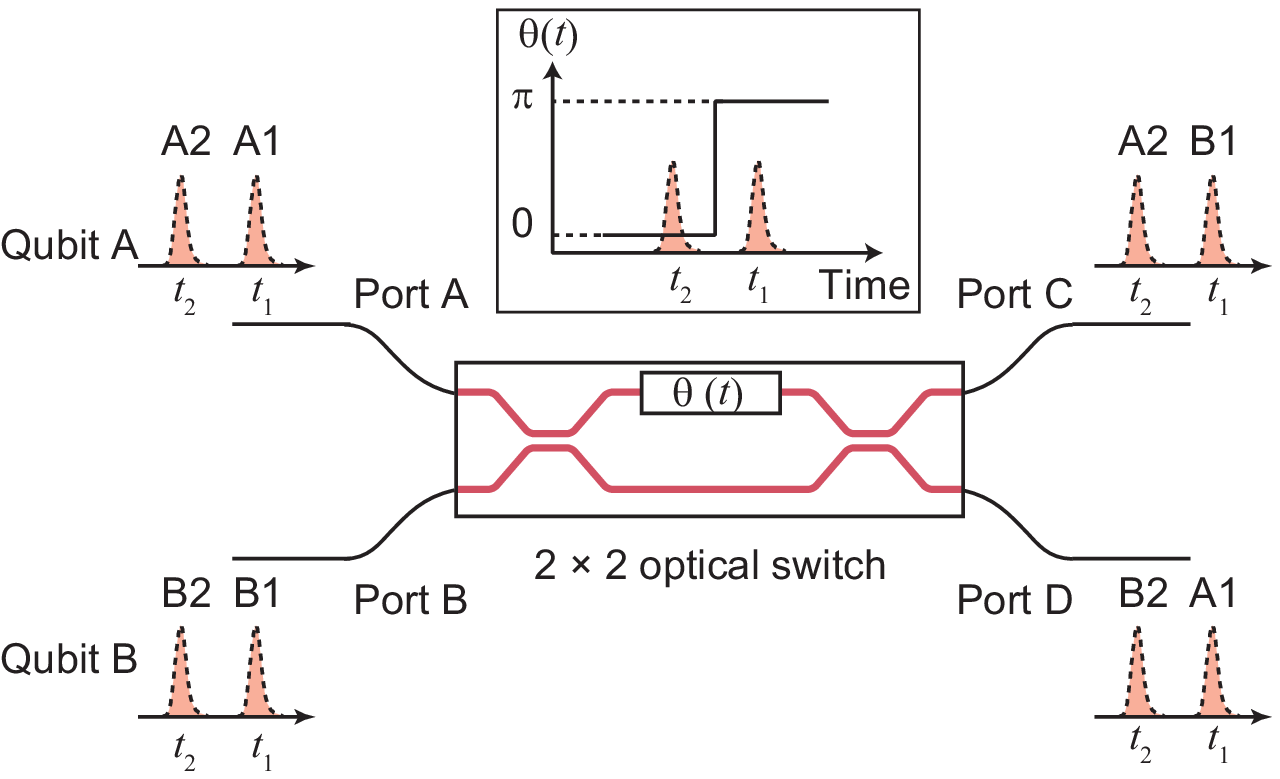}}

\caption{Concept of time-bin switching. Ax and Bx (x$=\{1,2\})$ correspond to the time bins in $t_1$ and $t_2$ at ports A and B, respectively. $\theta (t)$ is the phase difference between the two arms of the 2x2 optical switch in time $t$. 
}
\label{concept}

\end{figure}

Figure \ref{concept} shows the concept of a time-bin qubit entangler, which is based on two-photon quantum interference at a Mach-Zehnder interferometer optical switch with 2 input and 2 output ports (2x2 optical switch, hereafter). We launch two time-bin qubits whose states are given by $\frac{1}{\sqrt{2}} (|1_{t_1}\rangle_A + e^{i \phi_A} |1_{t_2}\rangle_A)$ and $\frac{1}{\sqrt{2}} (|1_{t_1}\rangle_B + e^{i \phi_B} |1_{t_2}\rangle_B)$ into ports A and B of the switch, respectively. Here, $|n_{t_x}\rangle_y$ denotes the state where there are $n$ photons in a temporal position $t_x$ and a spatial mode $y$. $t_1$ and $t_2$ correspond to the temporal positions of the first and second temporal modes that compose the time-bin qubits. The evolution of the states with the switch is described as
\begin{eqnarray}
|1_{t_k}\rangle_A &=& \cos \left(\frac{\theta (t_k)}{2}\right) |1_{t_k}\rangle_C - \sin \left(\frac{\theta (t_k)}{2}\right) |1_{t_k}\rangle_D, \label{eq1} \\
|1_{t_k}\rangle_B &=& \sin \left(\frac{\theta (t_k)}{2}\right) |1_{t_k}\rangle_C + \cos \left(\frac{\theta (t_k)}{2}\right) |1_{t_k}\rangle_D, \label{eq2}
\end{eqnarray}
where subscripts $C$ and $D$ show the modes at ports C and D, respectively, and $\theta (t_k)$ denotes the phase difference between the two arms of the Mach-Zehnder interferometer at time $t_k$. For the entangling operation, I set $\theta (t_1)=\pi$ and $\theta (t_2) =0$, as shown in the inset of Fig. \ref{concept}. This means that the spatial modes of the launched photons are exchanged at the first time slot and unchanged at the second time slot, as depicted in Fig. \ref{concept}. Note that this operation is equivalent to the operation of polarization modes using a polarization beam splitter. With this operation, and under the condition that the temporal and frequency distinguishability of the two time-bin qubits is eliminated, the state of the entire system is converted as follows. 
\begin{eqnarray}
& & \frac{1}{2} (|1_{t_1}\rangle_A + e^{i \phi_A} |1_{t_2}\rangle_A) \otimes (|1_{t_1}\rangle_B + e^{i \phi_B} |1_{t_2}\rangle_B) \nonumber \\
&\to& \frac{1}{2} (-|1_{t_1}\rangle_C|1_{t_1}\rangle_D -e^{i \phi_B} |0\rangle_C |1_{t_1}\rangle_D|1_{t_2}\rangle_D  + e^{i \phi_A} |1_{t_1}\rangle_C|1_{t_2}\rangle_C |0\rangle_D  + e^{i(\phi_A + \phi_B)} |1_{t_2}\rangle_C |1_{t_2}\rangle_D) \label{eq3}
\end{eqnarray}
The photons output from ports C and D are received by Charlie and David, respectively. By measuring the coincidences between Charlie and David, we can post-select a time-bin entanglement whose state is given by
\begin{equation}
\frac{1}{\sqrt{2}}(-|1_{t_1}\rangle_C|1_{t_1}\rangle_D + e^{i(\phi_A + \phi_B)} |1_{t_2}\rangle_C |1_{t_2}\rangle_D). \label{eq4}
\end{equation}

\begin{figure*}[thb]

\centerline{\includegraphics[width=\linewidth]{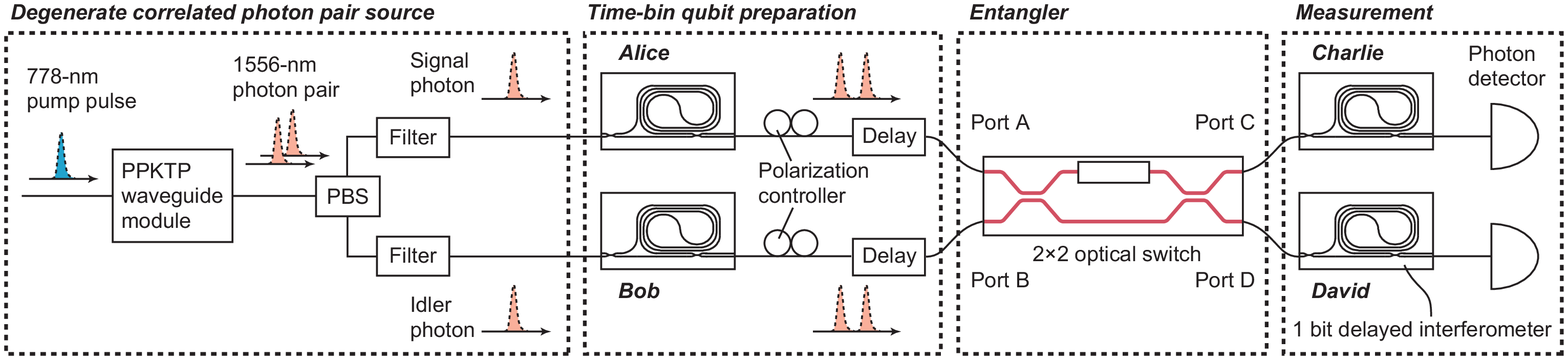}}

\caption{Experimental setup. }
\label{setup}

\end{figure*}

The proposed scheme was implemented using the setup shown in Fig. \ref{setup}. 
778-nm pulses were generated using second harmonic generation in a periodically poled lithium niobate waveguide pumped by 1556-nm pulses whose clock frequency and pulse width were 100 MHz and 60 ps, respectively. The 778-nm pulses were then launched into the fiber-coupled module of a periodically-poled potassium titanyl phosphate (PPKTP) waveguide (AdvR), where pulsed degenerate correlated photon pairs with a wavelength of 1556 nm were generated via type-II spontaneous parametric downconversion. The 778-nm pump power was adjusted so that the average number of correlated photon pairs per pulse became 0.25. The signal and idler photons were separated with a fiber-coupled PBS module, and then passed through optical bandpass filters with a 0.7-nm bandwidth. The signal and idler photons were then sent to Alice and Bob, respectively, who prepared time-bin qubits, by launching the photons into 1-bit delayed interferometers based on silica waveguides \cite{honjo,timebin}. After adjusting the temporal positions and the polarization states of the photons so that the temporal and polarization distinguishabilities of the two qubits were erased, the photons were input into ports A and B of a 2x2 optical switch based on a lithium niobate waveguide (EO Space), where the above time-bin switching operation took place. 
Here, the insertion loss of the switch was $\sim$4 dB, and the bandwidth and the extinction ratio of the switching were 10 GHz and 20 dB, respectively. 
The photons output from ports C and D were received by Charlie and David who were equipped with 1-bit delayed silica interferometers followed by single photon detectors based on InGaAs/InP avalanche photodiodes operated in a gated mode whose gate frequency was 100 MHz (Id Quantique). The detection efficiency and dark count rates for both detectors were 8\% and $2 \times 10^{-6}$ per gate, respectively. 
A state $|1_{t_x}\rangle_y$ is converted to $(|1_{t_x}\rangle_y + e^{i \phi_y} |1_{t_{x+1}}\rangle_y)/2$ with the 1-bit delayed interferometer, where $\phi_y$ denotes the phase difference between the two arms of the interferometer for mode $y$. 
Then, the state given by Eq. (\ref{eq4}) is converted to 
\begin{equation}
-|1_{t_1}\rangle_C|1_{t_1}\rangle_D + (e^{i (\phi_A + \phi_B)} - e^{i (\phi_C + \phi_D)}) |1_{t_2} \rangle_C |1_{t_2}\rangle_D
+ e^{i (\phi_A + \phi_B+\phi_C + \phi_D)} |1_{t_3} \rangle_C |1_{t_3} \rangle_D,
\end{equation}
where only the terms that contributed to the coincidence detection are shown and the normalizing term is discarded for simplicity. This equation indicates that we observe two-photon interference with the coincidence events at the 2nd time slot \cite{brendel,timebin}. Therefore, the detector gate positions were set at the slot. The detection signals from Charlie's and David's detectors were input into a time interval analyzer as the start and stop pulses, where coincidences were counted. The coincidence probability $P$ is given by
\begin{equation}
P \propto 1 - V\cos(\phi_A + \phi_B -\phi_C-\phi_D)
\end{equation}
where $V$ is the two-photon interference visibility.

\begin{figure}[thb]

\centerline{\includegraphics[width=.7\linewidth]{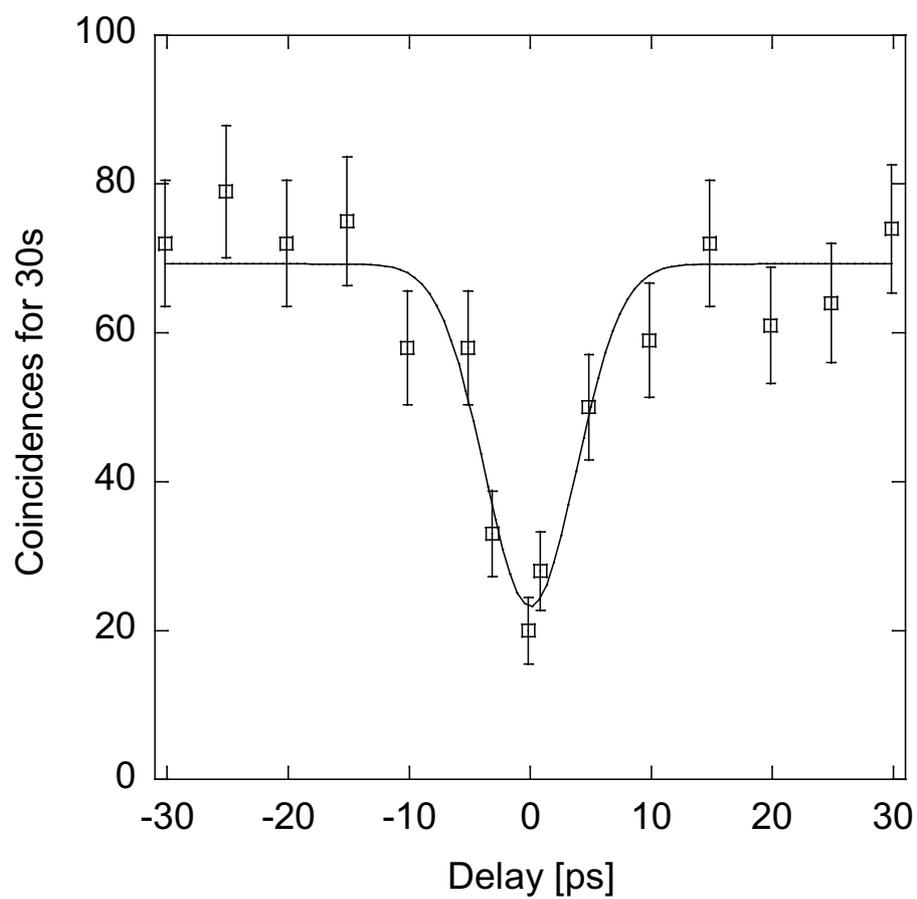}}

\caption{Hong-Ou-Mandel dip. Coincidences were recorded for 30 s and statistical error bars are shown. }
\label{hom}

\end{figure}

I first undertook a Hong-Ou-Mandel (HOM) quantum interference experiment \cite{hom} to confirm the indistinguishability of the photons from the degenerate pair source. I set $\theta (t_1)=\theta (t_2) = \pi/2$, which means that the 2x2 optical switch now works as a 50\% beamsplitter. I then removed the 1-bit delayed interferometers at Charlie's and David's sites, and measured the coincidences between the two detectors as a function of the relative delay between two photons. Here, the detector gate position was set at the first pulse. The result is shown in Fig. \ref{hom}. A clear dip in the coincidence counts is observed. The visibility of the dip was $66.4\pm 6.7$\%. Thus, a non-classical quantum interference was obtained.

\begin{figure}[thb]

\centerline{\includegraphics[width=.8\linewidth]{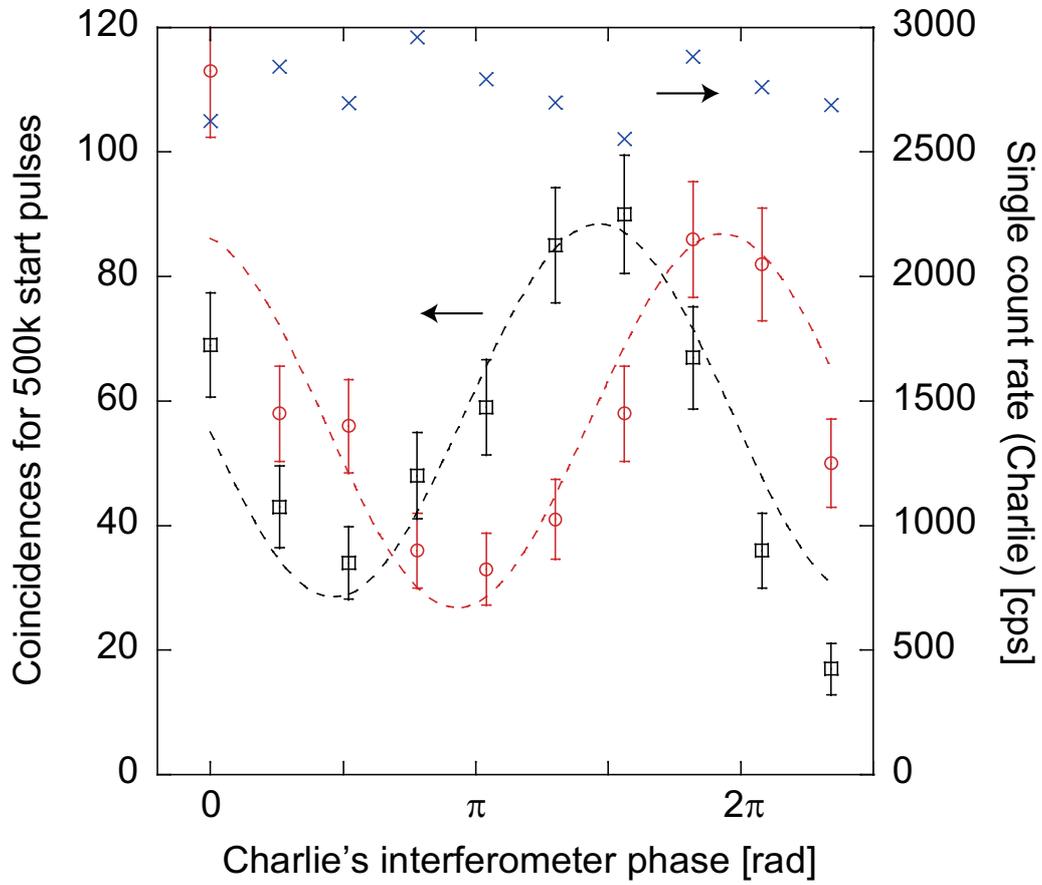}}

\caption{Coincidence fringes and single count rate as a function of Charlie's interferometer phase. The squares and circles show the coincidences when David's interferometer phase was set at 0 and $\pi/2$, respectively. The crosses represent the single count rate of Charlie's detector. The coincidences were obtained for 500,000 start pulses, and the statistical error bars are shown. }
\label{fringe}

\end{figure}

I then inserted Charlie's and David's interferometers into the setup, and measured the single count rate of Charlie's detector and the coincidences between the two detectors as a function of Charlie's interferometer phase. 
The result is shown in Fig. \ref{fringe}. The crosses show the single count rate, which was almost independent of Charlie's interferometer phase, suggesting that each photon was in a mixed state. On the other hand, the coincidences showed clear sinusoidal modulations for two non-orthogonal measurement bases at David. The raw visibilities of the fringes were $52.8\pm 12.9$\% for David's interferometer phase 0 and $51.2\pm 8.1$\% for $\pi/2$. 
If we assume a Werner state, a visibility greater than 33\% indicates the existence of entanglement \cite{peres}. Moreover, when the accidental coincidences were subtracted, the visibilities were $79.4\pm 20.6$\% (phase 0), and $84.4\pm 12.6$\% ($\pi/2$). Therefore, the obtained result suggests the successful generation of an entangled state using the proposed scheme. 

\begin{figure}[thb]

\centerline{\includegraphics[width=.7\linewidth]{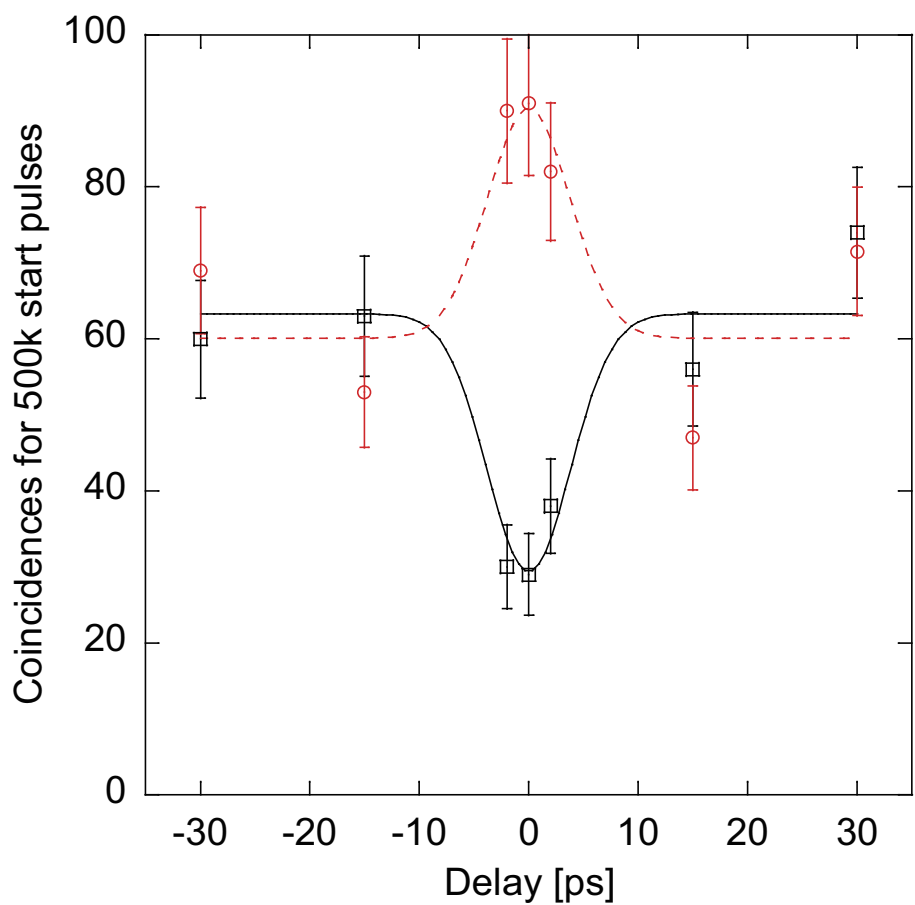}}

\caption{Coincidences as a function of the relative delay time between photons from Alice and Bob.  David's interferometer phase was set at 0, and Charlie's interferometer phase was set at $\pi$ (squares) and $2\pi$ (circles). The coincidences were obtained for 500,000 start pulses, and the statistical error bars are shown. }
\label{hop}

\end{figure}

To confirm that the fringes were generated as a result of quantum interference, I undertook another coincidence counting experiment, where I changed the relative delay between two photons while setting Charlie's interferometer phase at $2\pi$ and $\pi$, with David's phase at 0. The result is shown in Fig. \ref{hop}. Clear bunching (Charlie's phase $2\pi$) and anti-bunching ($\pi$) were observed only when the relative delay was set close to zero, which clearly indicates that HOM two-photon interference at the optical switch played a crucial role in the formation of entanglement.


Unfortunately, the obtained raw visibilities were not large enough to violate Bell's inequality. The main reason for the limited visibilities is the accidental coincidences caused by multi-photon emissions from the degenerate correlated photon pair source. Since I used a relatively large average photon-pair number per pulse of 0.25, the multi-photon emission probability was significantly enhanced. We can expect better visibilities simply by reducing the average photon number per pulse, at an expense of an extended measurement time. However, in the present experiment, the optimum bias voltage of the 2x2 optical switch drifted during the measurement, which degraded the switching extinction ratio. This temporal degradation of the extinction ratio limited the measurement time to about an hour, and prevented me from undertaking a longer measurement. Important future work will be to introduce feedback control of the bias voltage and so maintain a good switching extinction ratio for a longer period. This will allow us to undertake experiments with fewer accidental coincidences using a smaller average photon-pair number.


Since the function of time-bin switching is identical to that of a PBS, we can implement any quantum gates, whose functions were realized for polarization qubits using PBSs, for time-bin qubits using 2x2 optical switches. 
Moreover, Eqs. (\ref{eq1}) and (\ref{eq2}) indicate that the 2x2 optical switch can be used as a tunable time-bin switch. With this characteristic, we may implement various quantum gates for time-bin qubits. For example, by setting $\theta (t_1) = 2 \cos^{-1} (1/\sqrt{3})$ and $\theta (t_2) =0$, we can realize ``partial PBS" \cite{langford,okamoto} for time-bin qubits. Therefore, by a simple analogy with the CZ gate for polarization qubits \cite{langford}, we can implement a CZ gate for time-bin qubits by using three switches.

It is important to point out that Bell state measurements (BSM) of time bin encoded qubits using beamsplitters \cite{geneva,swp2,swp} can also be regarded as an entangling operation. In such a case, the two photons input into a beamsplitter are projected into the singlet state. However, unlike our scheme where the entangled photons can be used for further quantum gate operations, the photons in the BSM are destructively measured and so can not be used for future operations.


I expect that this scheme will be useful for advanced quantum communication experiments that involve quantum gates or cluster states over optical fiber networks \cite{azuma}. In addition, as with the switch used in the present experiment, high-speed 2x2 optical switches are usually realized using lithium niobate waveguides, which means that the proposed scheme is compatible with integrated quantum photonics. For example, the hybrid integration of high-speed optical switches based on lithium niobate waveguides and silica waveguides has been reported \cite{suzuki}. With those technologies, the proposed scheme may provide an important building block in sophisticated QIP systems based on integrated quantum photonics.

I thank T. Nakahara for lending me the 2x2 optical switch. 
I also acknowledge fruitful discussions with W. J. Munro, N. Matsuda, T. Inagaki, and K. Azuma.

\clearpage

\clearpage

\clearpage

\clearpage


\clearpage

\end{document}